# A Statistical Model of Information Evaporation of Perfectly Reflecting Black Holes

Laszlo Gyongyosi

[1] Quantum Technologies Laboratory, Department of Telecommunications
*Budapest University of Technology and Economics*
2 Magyar tudosok krt, Budapest, *H*-1117, Hungary
[2] MTA-BME Information Systems Research Group
*Hungarian Academy of Sciences*
7 Nador st., Budapest, *H*-1051, Hungary

gyongyosi@hit.bme.hu

**Abstract**

We provide a statistical communication model for the phenomenon of quantum information evaporation from black holes. A black hole behaves as a reflecting quantum channel in a very special regime, which allows for a receiver to perfectly recover the absorbed quantum information. The quantum channel of a perfectly reflecting (PR) black hole is the probabilistically weighted sum of infinitely many qubit cloning channels. In this work, we reveal the statistical communication background of the information evaporation process of PR black holes. We show that the density of the cloned quantum particles in function of the PR black hole's mass approximates a Chi-square distribution, while the stimulated emission process is characterized by zero-mean, circular symmetric complex Gaussian random variables. The results lead to the existence of Rayleigh random distributed coefficients in the probability density evolution, which confirms the presence of Rayleigh fading (a special type of random fluctuation) in the statistical communication model of black hole information evaporation.

# 1 Introduction

The process of quantum information transmission through a perfectly reflecting (PR) black hole (BH) can be evaluated via a noisy quantum channel that clones the absorbed quantum system and evaporates the cloned particles [1]. The application of quantum Shannon theory [14] in black hole physics [2-10], [19], [21-30], reduced the description of the evaporation process into the use of the tools of this field and made the situation well tractable. The black hole evaporation is still actively studied, particularly the appropriate statistical model for the internal evolution processes is still missing. In this work, we focus on the special regime, called the PR-regime (perfectly reflecting), in which a black hole works as a perfectly reflecting quantum channel, and we study the processing and evaporation of quantum information. In the description of the transmission process, we refer to a sender, Alice, who has fed in her quantum information into a PR black hole, and a receiver, Bob who wants to recover Alice's inputted quantum information. After Alice has thrown her quantum information into a PR black hole, it starts to process the absorbed quantum system and then evaporates the lower fidelity cloned quantum particles to Bob. The characterization of a PR black hole channel has been recently investigated in [1], and it has been found that the quantum channel of a PR black hole is the probabilistically weighted sum of infinitely many independent qubit cloning channels. However, there is no an appropriate statistical communication theory behind the flow of quantum-level communications of a PR black hole. Our aim is to give a statistical model for the information evaporation process of a PR black hole by exploiting the fundamental results of quantum Shannon theory and statistical communication theory.

In statistical communication theory, the analytical work is made by probabilistic models and well characterized mathematical background [11-13]. An important result from this field is the appropriate statistical description of signal propagation effects. In a noisy communication scenario, the transmit signal reaches to the receiver through a transmission medium, which causes attenuation and interference in the received signal. In the traditional statistical model, the input transmit signal is represented by an ideal Dirac pulse at time 0. The *fading* effect is a random variation (fluctuation) in the signal and it causes the degradation of the signal quality. In particular, under *Rayleigh fading*, the magnitudes of the signals that passed through a transmission medium follow Rayleigh random distribution [13], which represents an appropriate description of signal spread in wireless environments [20]. It is also a reasonable and a well-applicable wave propagation model if many small scatters are present in the environment, such as in dense urban areas or in tropospheric, ionospheric signal propagation.

Here, we show that the Rayleigh fading is also present in a PR black hole communication scenario, and it exists in a perfect symbiosis with the fundamental laws of quantum mechanics; however, the meaning is completely different in comparison to the traditional meaning of this phenomenon. What are the differences of the interpretation of fading effect in a traditional communication model and in the black hole communication scenario? Basically, while in traditional statistical communications the fading occurs on *physical signals*, in the proposed statistical communication model of black hole evaporation the fading effect will bring up on the *probability density* coefficients of the independent qubit cloning channel instances in the PR black hole. These coeffi-cients are Rayleigh random distributed and characterize the probability distribution of the cloning channels in the information evaporation process. This analogue between the traditional statistical communication theory and the black hole communication can be exploited further to reveal the statistical background behind the information evaporation of PR black holes, besides the fact that the probability density function is not a physical signal. Particularly, if Alice feeds in her quantum information at $t=0$, the noisy evolution of a PR black hole clones the input quantum particle via the phenomenon of *stimulated emission*, and then at $t>0$ it evaporates the resulting cloned particles to Bob (*Note*: Any matter or radiation fed into a PR black hole by Alice stimulates the emission of the cloned particles outside the event horizon [1], [10].). Since the cloning process at a PR black hole horizon cannot violate the no-cloning theorem, it follows that the fidelity of the evaporated clones will be lower than the fidelity of the absorbed input. The properties of a PR black hole channel are well characterized [1], as such, we do not include these results here. The details of the stimulated emission process are elaborated further in [10], [16-19].

Yet, the statistical background of the information evaporation is still unclear, and we do not have an appropriate statistical communication model that provides an answer to the internal evolution processes of a PR black hole. As a main purpose, we would like to reveal the statistical communication background of quantum information evaporation of black holes in the PR-regime.

This paper is organized as follows. In Section 2 the properties of the black hole quantum channel are summarized. Section 3 provides the theorems and proofs. The results are concluded in Section 4.

## 2 Quantum Channel of a PR Black Hole

First, we give a brief summary of the black hole channel focusing on the PR-regime and on the transmission of quantum information.

Assuming an input at $t=0$, the PR black hole is initialized with a nonzero mass indicator $z>0$, where $z$ is normalized onto the range of $0 \leq z < 1$. The mass indicator parameter is related to the bare black hole mass $m$; as $z \to 0$ the bare mass converges to zero, $m \to 0$, while for $z \to 1$ the bare mass converges to infinity, $m \to \infty$ (Note the utilization of parameter $z$ is required by the closed formulas of the PR-BH hole quantum channel $\mathcal{N}$.).

At $t>0$ decreases the probability of the ideal (noiseless) $1 \to 1$ channel realization to $p_1(t) = (1-z)^3$, while the $p_N$ probabilities of the noisy output realizations $1 \to N$, $N=2,\ldots,\infty$ (where $N$ stands for the number of output particles) monotonically increase with $z$. The black hole channel $\mathcal{N}$ in the PR-regime (PR-BH quantum channel) has the expression of

$$\mathcal{N} = \sum_{N=1}^{\infty} p_N \mathcal{Cl}_N, \qquad (1)$$

where $\mathcal{Cl}_N$ is the $1 \to N$ qubit cloning channel, while the $p_N$ probability of the $\mathcal{Cl}_N = 1 \to N$, $N=1,\ldots,\infty$ cloner is evaluated as $p_N = 1/2(1-z)^3(N+1)N^{(N-1)}$.

Feeding a quantum system $\psi$ into $\mathcal{Cl}_N$ results in

$$Cl_N(\psi) = \frac{N}{2}(N+1)\sum_{k=0}^{N} k|k\rangle\langle k|. \tag{2}$$

Note that $\psi$ in (2) can be an arbitrary qubit input, because the qubit cloning channel is $SU(2)$ covariant [15-16]. At a normalized black hole mass indicator $0 \leq z < 1$, $\mathcal{N}$ can be formulated as

$$\mathcal{N}(\rho) = (1-z)^3 \sum_{k=0}^{\infty} z^k \xi_k, \tag{3}$$

where $\rho = \frac{1}{2}(I + \hat{n} \cdot \vec{\sigma})$, $\vec{\sigma} = (\sigma_X, \sigma_Y, \sigma_Z)^T$, and $\sigma_X, \sigma_Y, \sigma_Z$ are the Pauli matrices, $\xi_k = \frac{(k+1)I^{k+2}}{2} + \hat{n} \cdot \vec{J}^{k+2}$ are defined in the $(k+2)$-dimensional space, while $\vec{J}^{k+2} = (\vec{J}_x^{k+2}, \vec{J}_y^{k+2}, \vec{J}_z^{k+2})$ are generators of the $(k+2)$-dimensional representation of $SU(2)$ [19]. For a maximally mixed input $\rho = \frac{1}{2}I$, one obtains output system $\sigma_B = \sum_{k=0}^{\infty} T_k S_k I^{k+2}$, and environment state $\sigma_E = \sum_{k=0}^{\infty} T_k \tilde{S}_k I^{k+1}$, where $T_k = (1-z)^3 z^k$, $S_k = \frac{1}{2}(k+1)$ and $\tilde{S}_k = \frac{1}{2}(k+2)$.

The PR black hole channel is conjugate degradable [1] and its quantum capacity is well tractable, $Q(\mathcal{N}) = \frac{(1-z)^3}{2} \sum_{k=0}^{\infty} (k+1)(k+2) z^k (\log_2(k+2) - \log_2(k+1))$. For a detailed analysis see [1] and for further background material, we suggest [16-19].

## 3 Theorems and Proofs

The main results are summarized in Theorems 1 and 2.

**Theorem 1** (Rayleigh random coefficients in the probability density evolution). *Let $p_N$ be the probability of $1 \to N$, $N = 1,...,\infty$ in the PR black hole, and let $\mathbf{c}_N(t) = \bigcup_{i=1}^{N} \sqrt{\Gamma_i(t)} \in \mathbb{R}^N$ be an N-dimensional coefficient vector, where $\sum_{i=1}^{N} \sqrt{\Gamma_i(t)} = \tau \sqrt{p_N}$, $\sqrt{\Gamma_i(t)} = |w_i(t)| \in \mathbb{R}$ is a Rayleigh random variable, $w_i(t) \in \mathcal{CN}(0, \sigma_{w_i}^2)$ is a zero-mean, circular symmetric complex Gaussian random variable with variance $\sigma_{w_i}^2 = \mathbb{E}[|w_i|^2]$, while $\tau$ is a normalization term. For $N = 1,...,\infty$, $p_N$ has the extension of $p_N = |\mathbf{c}_N(t)|^2 = \sum_{i=1}^{N} \Gamma_i(t) \in \mathbb{R}$, where $\Gamma_i(t) = |w_i(t)|^2 \in \mathbb{R}$ is exponentially distributed.*

*Proof.*
At $t = 0$, the mass indicator of a black hole is $z = 0$ and vector $\mathbf{c}_1(t=0) \in \mathbb{R}^N$ with magnitude $|\mathbf{c}_1(t=0)|$ identifies the $1 \to 1$ ideal map. At $t > 0$, as $z \to \infty$, the probability of $1 \to 1$ converges to zero, and the PR black hole evolutes the $N$-dimensional vector $\mathbf{c}_N(t) \in \mathbb{R}^N$ of the noisy $1 \rightarrow N$, $N = 2,...,\infty$ cloning channel instances. The $\mathbf{c}_N(t)$ coefficient vectors have a dot product $|\mathbf{c}_N(t)|^2 = \mathbf{c}_N(t) \cdot \mathbf{c}_N(t) \in \mathbb{R}$ that completely characterizes the output probabilities of the stimulated emission process in a PR black hole channel, as follows. First of all, to reveal the statistical communication background of the information evaporation we define some variables in the mathematical model.

Let $w_i(t)$ be a zero-mean, circular symmetric complex Gaussian random variable, $w_i(t) \in \mathcal{CN}(0, \sigma_{w_i}^2)$, with variance $\sigma_{w_i}^2 = \mathbb{E}[|w_i|^2]$. Then, let $a_j$ be a circular symmetric complex random phasor $a_j = r_j e^{i\phi_j} \in \mathbb{C}$, with $\mathrm{Re}(a_j) \in \mathbb{R}$, $\mathrm{Im}(a_j) \in \mathbb{R}$ i.i.d. random real and imaginary parts, and $\phi_j \in [0, 2\pi]$, $r_j \in \mathbb{R}$. For this complex random variable, $e^{i\varphi_i} a_j$ has the same distribution of $a_j$, $\mathbb{E}[a_j] = \mathbb{E}[e^{i\varphi_i} a_j] = \mathbb{E} e^{i\varphi_i}[a_j]$ for any $\varphi_i \in [0, 2\pi]$, by theory. A useful property of $w_i(t)$ that it can be used to model the squared magnitudes during the evaporation. This random variable can be rewritten as the sum of $j$ independent, circular symmetric complex random variables, for all $1 \rightarrow N$, and this connection is rooted in the following.

Let the sum of $j \rightarrow \infty$ independent circular symmetric complex random phasors be denoted as

$$\sum_j a_j = \sum_j r_j e^{i\phi_j}. \tag{4}$$

Then, from the *Central Limit Theorem* follows that (4), in fact, formulates a zero-mean, circular symmetric complex Gaussian random variable $w_i \in \mathcal{CN}(0, \sigma_{w_i}^2)$ with variance $\sigma_{w_i}^2 = \mathbb{E}[|w_i|^2]$, hence

$$\sum_j a_j = w_i = \mathrm{Re}(w_i) + i\,\mathrm{Im}(w_i) \in \mathbb{C}, \tag{5}$$

where $\mathrm{Re}(w_i) \in \mathcal{N}(0, 0.5\sigma_{w_i}^2)$, $\mathrm{Im}(w_i) \in \mathcal{N}(0, 0.5\sigma_{w_i}^2)$ are i.i.d. zero-mean Gaussian random real and imaginary parts of $w_i \in \mathcal{CN}(0, \sigma_{w_i}^2)$. From the complex circular symmetry property follows that $e^{i\varphi_i} w_i$ has the same distribution of $w_i$, $\mathbb{E}[w_i] = \mathbb{E}[e^{i\varphi_i} w_i] = \mathbb{E} e^{i\varphi_i}[w_i]$, for any $\varphi_i \in [0, 2\pi]$. The decomposition shown in (5) can be extended for all $w_i$ components of the sum $\sum_{i=1}^{N} |w_i(t)|$, $i = 1...N$. Taking the magnitude of $w_i(t)$ leads to precisely

$$\begin{aligned}|w_i(t)| &= \left|\sum_j a_j\right| \\ &= \left|\sum_j r_j e^{i\phi_j}\right| \\ &= \sqrt{\mathrm{Re}(w_i)^2 + \mathrm{Im}(w_i)^2} \\ &= \sqrt{\Gamma_i(t)}.\end{aligned} \tag{6}$$

The variable $\sqrt{\Gamma_i(t)} \in \mathbb{R}$ is a Rayleigh random variable with density

$$f(|w_i(t)|) = \left(|w_i(t)|/\sigma_{w_i}^2\right) e^{-|w_i(t)|^2/2\sigma_{w_i}^2}, \quad |w_i(t)| \geq 0. \tag{7}$$

The squared magnitude $\Gamma_i(t) = |w_i(t)|^2 \in \mathbb{R}$ is exponentially distributed with density

$$f(|w_i(t)|^2) = \left(1/\sigma_{w_i}^2\right) e^{-|w_i(t)|^2/\sigma_{w_i}^2}, \quad |w_i(t)|^2 \geq 0. \tag{8}$$

Second, the black hole channel in the PR-regime has the special *orthogonal convex sum structure* (direct sum, see (1)), and the total output probability density equals to the direct sum of

$$\begin{aligned} p_{tot} &= \sum_{N=1}^{\infty} p_N \\ &= \sum_{N=1}^{\infty} |\mathbf{c}_N(t)|^2, \end{aligned} \tag{9}$$

where $p_N$ is the probability of the realization of the $1 \to N$ cloning channel in the orthogonal convex sum channel. The variables $w_i \in \mathcal{CN}(0, \sigma_{w_i}^2)$-s are zero-mean, circular symmetric complex Gaussian random variables for all $i$, $i = 1 \ldots N$, and for all $1 \to N$, $N = 1, \ldots, \infty$, it follows that the magnitudes of $w_i$-s, $|w_i(t)| = \sqrt{\Gamma_i(t)}$ are Rayleigh random variables, by theory. It immediately indicates that the sum of the elements of $\mathbf{c}_N(t)$ for $1 \to N$, $N = 1, \ldots, \infty$, in fact, can be expressed in terms of $\sqrt{\Gamma_i}$ as $\sqrt{p_N} = \frac{1}{\tau} \sum_{i=1}^{N} \sqrt{\Gamma_i(t)}$.

The cross-verification of these statements requires the detailed analysis of the output probability density that will be presented in the proof of Theorem 2, however these results already demonstrate that the stimulated emission process of a PR black hole is related to Rayleigh fading by the Rayleigh random magnitude coefficients $\sqrt{\Gamma_i(t)} = |w_i(t)|$ of the independent cloning channels $1 \to N$, $N = 1, \ldots, \infty$. Similarly, the squared magnitudes, $\Gamma_i(t) = |w_i(t)|^2$, are exponentially distributed variables, see (8).

The existence of zero-mean, circular symmetric complex Gaussian random variables $w_i$ for all output realizations $1 \to N$, confirms the Rayleigh random distributed coefficients in the extension.

These results immediately prove the presence of Rayleigh fading in the statistical communication model of the information evaporation of a PR black hole.

∎

**Corollary 1**. *The total probability $p_{tot}$ of the output realizations $1 \to N$, $N = 1, \ldots, \infty$ in the stimulated emission of a PR black hole is as follows*

$$\begin{aligned}p_{tot} &= \sum_{N=1}^{\infty}\left|\mathbf{c}_N\left(t\right)\right|^2\\ &= \sum_{N=1}^{\infty}\sum_{i=1}^{N}\Gamma_i\left(t\right)\\ &= \sum_{N=1}^{\infty}\sum_{i=1}^{N}\left|w_i\left(t\right)\right|^2 \\ &= \sum_{N=1}^{\infty}\sum_{i=1}^{N}\left|\sum_j a_j\right|^2\\ &= 1,\end{aligned} \quad (10)$$

where

$$\sum_{i=1}^{N}\Gamma_i\left(t\right) = p_N, \quad (11)$$

$$\sqrt{\Gamma_i\left(t\right)} = \left|\sum_j a_j\right|, \quad (12)$$

and

$$\Gamma_i\left(t\right) = \left|w_i\left(t\right)\right|^2 = \left|\sum_j a_j\right|^2. \quad (13)$$

*Proof.*
These results trivially follow from the proof of Theorem 1; hence, the proof is immediately concluded here.

∎

**Theorem 2** (The distribution of the cloned particles in the information evaporation of PR black holes). *The $p_N$ probability density of $1 \to N$, $N = 1,...,\infty$ is $\chi^2_{2N}$ Chi-square distributed with 2N degrees of freedom. At a mass indicator z, it can be approximated by the density function $f_N\left(z\right) \approx \frac{1}{(N-1)!}\left(10z\right)^{N-1}e^{-10z}$.*

*Proof.*
First, from Theorem 1 and Corollary 1, we express $p_{tot}$ as

$$p_{tot} = \sum_{N=1}^{\infty}\sum_{i=1}^{N}\left|w_i\left(t\right)\right|^2. \quad (14)$$

Then we exploit that in the traditional statistical model of Rayleigh fading, the summand variables that characterize the magnitudes of the transmit signals are the magnitudes of zero-mean, circular symmetric complex Gaussian random variables (these complex variables are denoted by $w_i\left(t\right)$ in our scenario).

As one can readily see, in the proposed statistical model these variables bring up in the next sum:

$$p_N = \sum_{i=1}^{N} \Gamma_i(t) \\ = \sum_{i=1}^{N} |w_i(t)|^2. \qquad (15)$$

As follows, $p_N$ is, in fact, the sum of $N$ independent squared Rayleigh random variables. In other words, (15) is the sum of the squares of $2N$ independent, real, zero-mean Gaussian random variables.

According to Theorem 1, term $\Gamma_i(t)$ precisely equals to $\Gamma_i(t) = \text{Re}(w_i)^2 + \text{Im}(w_i)^2$, where $\text{Re}(w_i), \text{Im}(w_i) \in \mathbb{N}(0, 0.5\sigma_{w_i}^2)$, which clearly shows that the $p_N$ distribution of the $1 \to N$ qubit cloner instances, in function of $z$ of the PR black hole, can be characterized by a $f_N(z) \in \chi_{2N}^2$ Chi-square distribution with $2N$ degrees of freedom.

In particular, the density function $f_N(z)$ is an approximation of $\chi_{2N}^2$, where $2N$ arises from the fact that $w_i$-s are circular symmetric complex Gaussian random variables, with i.i.d $\mathbb{N}(0, 0.5\sigma_{w_i}^2)$ real and imaginary parts, respectively.

In function of $z$, the density of $p_N$ of $1 \to N$, $N = 1, \ldots, \infty$, can be evaluated by $f_N(z)$ as follows

$$f_N(z) \approx \frac{1}{(N-1)!}(10z)^{N-1} e^{-10z}. \qquad (16)$$

The $f_N(z)$ density of the realization of the $1 \to N$ qubit cloner instances in function of $z$ (for $N = 1, 2, 3, 4, 5$) in a PR black hole is depicted in Fig. 1.

Note, that the density function picks up a value of $f_N(z) < y$ with probability

$$\Pr[f_N(z) < y] \approx \int_0^y \frac{1}{(N-1)!} x^{N-1} dx = \frac{1}{N!} y^N, \qquad (17)$$

where $x = 10z$.

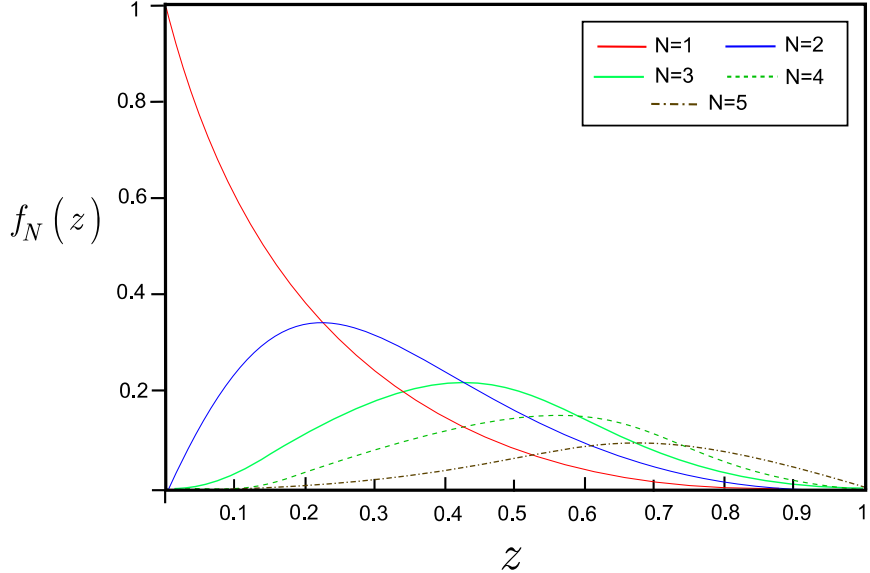

**Figure 1.** The density of $p_N$ for different values of $N$ in function of the mass indicator $z$ of a PR black hole. The function approximates a $\chi^2_{2N}$ Chi-square distribution with $2N$ degrees of freedom.

These results verify the statements of Theorem 1 and conclude the proof of Theorem 2.

∎

## 3.1 Brief overview

The proposed statistical communication model for the information evaporation process of a PR black hole is briefly sketched as follows. The fed in quantum system $\psi$ stimulates the emission of the lower fidelity clones at the event horizon, which is modeled by the $1 \to N$ cloning channels. A PR black hole is characterized by its mass indicator $z$, which determines the probability of the realization of an $1 \to N$ instance in the stimulated emission process.

At a given $z$, the output realizations have density $f_N(z)$ that can be approximated by a $\chi^2_{2N}$ Chi-square distribution with $2N$ degrees of freedom. The channel output $1 \to N$, $N = 1,\ldots,\infty$ has probability $p_N$, and it equals to $p_N = \sum_{i=1}^{N} \Gamma_i(t) = \sum_{i=1}^{N} |w_i(t)|^2$, while the total density is $p_{tot} = \sum_{N=1}^{\infty} \sum_{i=1}^{N} \Gamma_i(t) = \sum_{N=1}^{\infty} \sum_{i=1}^{N} |w_i(t)|^2 = \sum_{N=1}^{\infty} \sum_{i=1}^{N} \left| \sum_j a_j \right|^2$, where $w_i(t)$ is a zero-mean, circular symmetric complex Gaussian random variable $w_i(t) \in \mathcal{CN}\left(0, \sigma^2_{w_i}\right)$.

## 4 Conclusions

In this work, we proposed a statistical communication model for the information evaporation process of perfectly reflecting black holes, focusing on the transmission of quantum information.

As we have found, the statistical communication model of the probability density evolution in a PR black hole is connected to the statistical model of Rayleigh fading. In particular, we revealed that in the PR-regime the density of the evaporated clone particles follows Chi-square distribution, and the probabilities of the independent cloning channels are the sum of the squared magnitudes of independent complex Gaussian random variables. We showed that the stimulated emission is characterized by Rayleigh random variables in the orthogonal convex sum structure of a PR black hole channel.

# Acknowledgements

The results are supported by the grant COST Action MP1006.